\documentstyle[art12]{article}
\begin{document}

\begin{titlepage}
\begin{flushright}
FERMILAB-Pub-96/420-T\\
hep-ph/9611315\\
{\small November 1996}
\end{flushright}
\vspace{1cm}
\begin{center}
{\Large\bf Hara's Theorem and $W-$exchange   \\
in Hyperon Weak Radiative Decays}\\
\vspace{1cm}

{\large Ya.~Azimov}\\
\vspace{0.4cm}
{\it
Fermi National Accelerator Laboratory, P.~O.~Box 500,\\
Batavia, IL 60510, U.S.A.}\\
\vspace{0.4cm}
and\\
\vspace{0.4cm}
{\it 
Petersburg Nuclear Physics Institute,\\
Gatchina, St.Petersburg, 188350, Russia}\\
\vspace{0.4cm}
\vspace{0.4cm}
\end{center}

\begin{abstract}
We reconsider  Hara's theorem in its relation 
to the well-known properties of $\beta-$decay. 
All assumptions  necessary for the theorem to 
be true are explicitly formulated. Further, we 
study the $W-$exchange contribution to weak 
radiative decays and show that it does not 
violate Hara's theorem. However, this contribution 
reveals the essential role of particle mixing in 
symmetry considerations and some peculiar features 
of gauge-invariant amplitudes under perturbative 
expansion. Together they explain an effect, which 
was treated as contradicting Hara's theorem,
without any violation. The properties of 
$W-$exchange we describe here may have more 
general importance and should be taken into 
account in further detailed calculations of 
weak processes. \\
\vspace{0.4cm}

PACS numbers: 11.10.Gh, 12.15.Lk, 14.20.-c

\end{abstract}

\end{titlepage}

\section{Introduction}

It is a common belief that weak and electromagnetic 
interactions may always be treated perturbatively, 
generally in contrast to  strong interactions (SI). 
This makes many weak or electromagnetic processes 
involving hadrons a good arena for analyzing 
details of (nonperturbative) SI. They continue to 
provide a lot of phenomenological information on 
SI and present a vast testing ground for their 
theoretical description.

A special role could be played here by weak radiative 
decays (WRD) of hadrons. They exemplify rather rare 
hadron processes which are clearly of higher order 
in perturbations.  Therefore, their interplay 
with SI might give further insight into details of 
nonperturbative SI. Moreover, experimentally available 
WRD have quite simple two-body kinematics, very similar 
to weak nonleptonic decays. It is therefore surprising 
that, up to now, there is no generally acceptable 
theoretical description of WRD (see the recent~\cite{JLPZ} 
and older~\cite{YaA} reviews and references therein; see 
also recent talks~\cite{ Sin},\,\,\cite{PZ}).

The last decade has witnessed significant experimental 
progress that has produced a large set of data on WRD, 
mainly from CERN and FNAL hyperon beams (see, e.g., 
summary table in ref.~\cite{Tim}). Though still 
incomplete, the data give good evidence for an essential 
role of $W-$exchange in inducing WRD. Those decays in 
which a $W-$boson can be exchanged between valence 
quarks are at least an order of magnitude  more copious 
than decays that do not allow such an exchange.

This fact revived the old problem of the so-called 
Hara's theorem~\cite{Har}. According to the theorem, 
the exact  $SU(3)$ symmetry of SI would make some decays 
(in particular, $\Sigma^+ \rightarrow p\gamma$) have 
vanishing parity-violating amplitudes and, thus,  
vanishing decay asymmetries. Of course, the $SU(3)$ 
symmetry is violated in nature, so the deviation of 
experimental asymmetry~\cite{Tim} for $\Sigma^+ 
\rightarrow p\gamma$ from the prediction of Hara's 
theorem could arise quite naturally. But the large 
value of the asymmetry and its unexpected sign 
continue to be a hard problem of WRD theory.

The absence of a satisfactory simultaneous description  
for the $\Sigma^+ \rightarrow p\gamma$  decay width 
and asymmetry, together with the rather complicated 
character of the original proof, has, again and again, 
stimulated  skepticism as to the correctness  of Hara's 
theorem.  As an example, we recall ref.~\cite{LL}, 
that was opposed by  Gaillard~\cite{MG}. Now that 
experiment suggests the important role of 
$W-$exchange, the essential contribution to such 
skepticism comes from the paper of Kamal and 
Riazuddin~\cite{KR}.  They argue that explicit 
calculations of $W-$exchange at the quark level  
directly contradict Hara's theorem, even for exact 
$SU(3)$ symmetry. So the question as to why 
the asymmetry for $\Sigma^+\rightarrow p\gamma$ 
strongly deviates from the simple Hara's prediction 
becomes again topical~\cite{Sin}, \cite{PZ}. Its 
solution has been even assumed to be closely 
related to the general problem of correspondence 
between the quark and hadron levels~\cite{PZ}.

In the present note we reconsider the above problems. 
First we repeat in more detail Gaillard's arguments~
\cite{MG} for  correctness of Hara's theorem and its 
similarity to usual $\beta-$decay. We formulate 
explicitly all the assumptions necessary to prove 
the theorem.

Then, in the case of exact $U-$symmetry, we consider 
structure and properties of quark-  and hadron-level 
$W-$exchange contributions, both by themselves and 
accompanied by a photonic transition. On one side, 
we again prove Hara's theorem, for this particular 
case. On the other side, we demonstrate some simple, 
but unfamiliar properties of gauge-invariant  
amplitudes being expanded as perturbation series 
in weak interactions. These properties provide us 
the basis to understand the effect observed by Kamal 
and Riazuddin~\cite{KR} without invoking a violation 
of Hara's theorem. An interesting point of our 
consideration is the essential role of quark and 
hadron mixing which may be also important for  more 
detailed study of all WRD's or, even more generally,
of higher order corrections in weak interactions.

\section{$\beta-$decay and $\Sigma^+\rightarrow p\gamma$}

We begin with the usual $\beta-$decay $n \rightarrow 
pe\overline\nu$.  Its Lagrangian is proportional to 
\begin{equation}                                   
L_{\beta} \propto J^{(\beta)}_{\mu}j^{\mu}
+ J^{(\beta)\dagger}_{\mu}j^{\mu\dagger}, 
\end{equation}
where $j_{\mu}$  is the lepton weak charged current 
and $J^{(\beta)}_{\mu}$ is the nucleon weak charged 
current; note that here we are not interested in  the 
exact form of constant factors. It is well known that  
$J^{(\beta)}_{\mu}$ consists of two parts, the vector 
current  $J^{(\beta)V}_{\mu}$ and the axial one  
$J^{(\beta)A}_{\mu}$. Their matrix elements between 
neutron and proton have the additional structure 
\begin{equation}
<p|J^{(\beta)V}_{\mu}|n> = \overline\psi_p 
\left[f_1(k^2)\gamma_{\mu} +
f_2(k^2)\sigma_{\mu\nu}k^{\nu} 
+ f_3(k^2)k_{\mu}\right] \psi_n  ,
\end{equation}
\begin{equation}
<p|J^{(\beta)A}_{\mu}|n> = \overline\psi_p 
\left[g_1(k^2)\gamma_{\mu}\gamma_5 
+ g_2(k^2) i\sigma_{\mu\nu}k^{\nu} \gamma_5
+ g_3(k^2) i\gamma_5 k_{\mu}\right]\psi_n.
\end{equation}
Here $k$ is the momentum transfer and functions 
$f_i, g_i$ are real at $k^2$
below hadron production thresholds, e.g., at 
space-like $k$. For the Dirac matrices, we use the 
notations of ref.~\cite{BD}. 

Let us assume that isotopic symmetry is exact. Then 
every term of eqs.~(2),(3) is a component of an 
isotopic vector. Due to the fact that $p$ and $n$ are 
members of the same isotopic doublet, all isotopic 
vectors involved in eqs.~(2),(3) transform into 
themselves under $G-$transformations~\cite{LY} and
have some definite $G-$parity. Terms with $f_1, 
f_2, g_2 $ are $G-$even, and ones with $g_1, g_3, 
f_3 $ are $G-$odd. Weinberg~\cite{Wein} suggested
further to separate hadronic weak charged currents 
in strangeness-conserving processes into two classes. 
Then terms with $f_1, f_2, g_1, g_3$ belong to the first 
class, and ones with $f_3, g_2$ are of the second class. 
From $G-$parity conservation by strong interactions, 
Weinberg concluded~\cite{Wein} that first-class bare 
interactions can induce only first-class phenomenological
currents. As we know today strangeness-conserving hadron 
weak currents in the Standard Model are indeed of the first 
class. Therefore, axial weak magnetism (the term with 
$\sigma_{\mu\nu}\gamma_5 $) is forbidden in $\beta-$decay
(together with the induced scalar term $f_3 k_{\mu}$) 
while vector weak magnetism (the term $f_2\sigma_{\mu\nu}
k^{\nu}$) is permitted~\cite{G-M} (together with the 
induced pseudoscalar term $g_3 i\gamma_5k_{\mu}$)
and observable. 

Note that in $\beta$-decay, all the first-class currents  
have the same, and negative, $GP-$parity, while the 
$GP-$parity of the second-class terms would be positive. 
This allows one to relate absence of axial magnetism
directly to $CP-$conservation in $\beta-$decay, without
an independent assumption of absence of the second-class
currents. Indeed, the lepton weak currents $j_{\mu}$ and 
$j_{\mu}^{\dagger}$ are known to transform into each other 
by $CP$ transformation. So, if the Lagrangian~(1) is to be 
$CP-$conserving the hadron weak currents $J_{\mu}$ and 
$J_{\mu}^{\dagger}$ should also be related to each other 
by $CP-$transformation. An essential additional feature 
which comes from exact isospin symmetry is that the two 
currents are various combinations of components 
of the same isotopic vector. Moreover, they may be 
transformed into each other by isospin rotation 
around the 2nd axis, without changing coefficient 
functions in eqs.(2),(3). These two properties 
together lead just to the necessity of negative $GP-$parity.  
The opposite sign between the $CP-$parity of the Lagrangian 
(1) and the $GP-$parity of the currents arises from the 
rotation of isotopic-vector components involved into 
$G-$transformation.
 
Of course, all the above is true only up to electromagnetic 
radiative corrections (or, more exactly, up to violation of 
isotopic symmetry and, therefore, $G-$parity conservation). 
Thus, the axial magnetism is not forbidden at the level of 
radiative corrections.

Now we turn to the $\Sigma^+p\gamma$ interaction. We will
follow the same logic as briefly presented by Gaillard~\cite{MG}. 
In particular, we will use not the whole $SU(3)$ symmetry 
but a more narrow group, $U-$spin symmetry (the analog of 
isotopic $I-$spin symmetry which mixes $d,s$ quarks instead 
of $u,d$), which is known~\cite{Gour} to be sufficient for 
Hara's theorem.
 
The effective Lagrangian responsible for the $\Sigma^+p
\gamma$ interaction has the form 
\begin{equation}
L_S^{(\gamma)} \propto \left( J^{(S)}_{\mu} + 
  J^{(S)\dagger}_{\mu} \right) A^{\mu}
\end{equation}
which recalls eq.~(1). Here $A_{\mu}$ is the photon field.   
Currents  $J^{(S)}_{\mu}$ and $J^{(S)\dagger}_{\mu}$ 
satisfy strangeness selection rules $\Delta S = \pm 1$
respectively.  

If the $U-$spin symmetry of strong and electromagnetic 
interactions is exact we may introduce the $G_u-$transformation
analogous to the $G-$transformation of the usual isotopic group.
Now we can simply modify the previous consideration of 
$\beta-$decay and apply it to WRD's.

Again, the current $J^{(S)}_{\mu}$ has vector and axial parts.
Their  matrix elements are
\begin{equation}
<p|J^{(S)V}_{\mu}|\Sigma^+> = 
\overline\psi_p\left[f_1^{(S)}(k^2)\gamma_{\mu}
+ f^{(S)}_2(k^2)\sigma_{\mu\nu}k^{\nu} + 
f^{(S)}_3(k^2)k_{\mu}\right] \psi_{\Sigma^+} ,
\end{equation}
\begin{equation}
<p|J^{(S)A}_{\mu}|\Sigma^+> =  
\overline\psi_p\left[g^{(S)}_1(k^2)\gamma_{\mu}\gamma_5
+ g^{(S)}_2(k^2) i\sigma_{\mu\nu}\gamma_5k^{\nu} +
g^{(S)}_3(k^2) i\gamma_5k_{\mu}\right]\psi_{\Sigma^+} .
\end{equation}
An important point is that $p$ and $\Sigma^+$ are members 
of the same $U-$spin doublet, just as $p$ and $n$ belong to
the same $I-$spin doublet. In full similarity to the above
$\beta-$decay considerations, currents $J^{(S)}_{\mu}$  and
$J^{(S)\dagger}_{\mu}$ are related to each other  by both
Hermitian conjugation and $CP-$transformation. Exact $U-$spin
symmetry gives possibility to connect $CP-$transformation of 
$J^{(S)}_{\mu}$ into $J^{(S)\dagger}_{\mu}$ with 
$G_uP-$transformation of $J^{(S)}_{\mu}$  into itself.
Then we arrive at the result of Hara~\cite{Har} which
arises exactly as the above statements on $\beta-$decay. 

Now we can generalize and collect together necessary 
requirements for Hara's theorem:

1. Exact $U-$spin symmetry for strong and electromagnetic
interactions (violated, of course, by weak interactions).

2. Hermiticity of the effective Lagrangian for WRD. 

3. $CP-$conservation.

4. Initial and final hadrons in a particular WRD being 
members of the same $U-$spin multiplet.

5. Weak interactions having a structure that produces 
a nonvanishing transition amplitude for the WRD, 
with a unique $U-$spin structure (vector in the case of 
$\Sigma^+ \rightarrow p\gamma$).

The meaning of the first 3 conditions is obvious. The 
last two conditions guarantee a~definite $G_uP-$parity 
for all parts of the current $J^{(S)}_{\mu}$. They can 
be further generalized so to admit the transition 
amplitude being a mixture of various $U-$spin 
representations. The only problem is that all the 
involved representations should have odd integer $U-$spin 
to provide the same, negative, $G_uP-$parity.
  
Let us consider some particular cases. The pair  
($p\Sigma^+$) surely satisfies conditions 4 and 5 as they 
are. Its both members belong to the same $U-$spin doublet, 
and the transition amplitude can be only an $U-$vector. The 
same is true for another pair, ($\Sigma^-\Xi^-$). Thus, 
these pairs correspond to two decays where Hara's theorem 
could be applicable. 

On the other side, transition amplitude, e.g., in the 
decay  $\Lambda \rightarrow n\gamma$, have more 
complicated $U-$spin structure. Even $\Lambda$ itself 
is the mixture of $U-$singlet and $U-$vector components, 
while $n$ is a pure $U-$vector. So the current can have 
several parts, with various properties under $U-$spin 
rotations and various $G_uP-$parities. We see that 
conditions 4 and 5 are violated (even in a more 
generalized form; there are $U-$vector and $U-$tensor 
parts in the amplitude). As a result, Hara's theorem 
is definitely inapplicable here.   

The above conditions are simple and very general. 
Except $U-$symmetry, they are surely respected by 
many approaches that have been used to describe WRD. 
In particular, these assumptions are true for the 
$W-$exchange considered in ref.~\cite{KR}. 

Up to now our consideration has been rather 
formal and could not be applied to 
experiment. The necessary next step would be 
using gauge invariance. For the real photon 
($k^2 = 0$) it should eliminate the first terms 
in expressions (5), (6). The last terms
give no contribution when multiplied by the 
real-photon polarization vector. Then the vector 
and axial magnetic terms become the only physical 
terms, and Hara's theorem becomes operative for 
experiment. But here we postpone this step. We 
return to it after the study of $W-$exchange.

\section{Structure of the $W-$exchange contribution }

To understand in more detail some specific features of 
$W-$exchange in WRD, we begin by considering this
exchange by itself, without  any photon emission.
  
The $W-$exchange for the transition
$\Sigma^+ \rightarrow p$ corresponds to an interaction 
of the form
$$ (\overline u_i O s_i)[W](\overline d_j O u_j) $$ 
with summation over color indices $i, j$.
Here we explicitly show which quark pairs exchange
the $W-$boson just to emphasize the color structure of 
their  interaction. The coefficient contains product 
of the corresponding elements of the CKM-matrix.
Also possible are interactions
$$(\overline u_i O d_i)[W](\overline d_j O u_j)\,\,\,
{\rm and}\,\,\,
(\overline u_i O s_i)[W](\overline d_j O s_j) ,$$
where the hadrons may stay untransformed.

If $U-$symmetry is exact the masses of $d$ and $s$ 
quarks coincide. The same is true for $p$ and 
 $\Sigma^+$. Hence, we may mix these degenerate quarks 
(and hadrons). It is convenient for our purpose to choose 
the mixed quarks $d'$ and $s'$ so as to eliminate the 
transition  $s' \rightarrow u$  in the CKM-matrix  (note 
that in the case of 3 or more generations this does not 
mean vanishing of the relative transition $d'\rightarrow c$).  
Then the interaction $$ (\overline u_i O s'_i)[W](\overline 
d'_j O u_j)$$ becomes impossible and $\Sigma'^+$ can not 
be transformed to $p'$ by single $W-$exchange.

Now the only possible interaction between light quarks
due to $W-$exchange is
\begin{equation}
L_W \propto (\overline u_i O d'_i)[W](\overline d'_j O u_j).
\end{equation}
It interchanges quark flavors without permutation of 
colors. So the question arises whether interaction~(7) 
is capable of transforming $p'$ into itself without any 
excitation. The answer is yes, due to colorlessness 
of hadrons. Indeed, the color wave function of baryons is
totally antisymmetric, and color permutation may only
reverse the sign of transition amplitudes. For the standard 
$(V-A)$ vertices in the limit $m_W \rightarrow \infty$ this 
sign reversal may be eliminated by  Fierz transformation.
Then the $W-$exchange interaction (7) efficiently becomes  
totally symmetric under transposition of $u$ and $d'$ 
(or $\overline u$ and $\overline d'$). This point-like 
4-quark interaction can be effectively rewritten in the 
diagonal form $$ (\overline d'_i O d'_i)(\overline u_j O 
u_j) ,$$ where initial quarks retain their quantum numbers.
Such interaction is surely able to provide diagonal 
transition $p'\rightarrow p'$.

Thus, we have achieved diagonalization of the degenerate 
hadronic states  $p'$ and $\Sigma'^+$,
when no transition  between them (through $W-$exchange)
is possible, though they both can transform into heavier 
states which contain, e.g., $c$ or $b$ quarks. Moreover,
the interaction $$(\overline u_i O s'_i)[W](\overline s'_j 
O u_j)$$ also becomes impossible, so that $W-$exchange can 
not transform $\Sigma'^+$ into itself, while being able to 
transform $p'$ into itself.

Consider, for comparison, neutral members of the baryon 
octet $n,\,\Lambda,\,\Sigma^0,\,\Xi^0$. If the $U-$spin 
symmetry were initially exact, it would be useful to  
consider states $n'= (d'd'u)$, $\Xi'^0 =(s's'u)$ and two 
states $\Sigma'^0,\,\Lambda'$ which have the same quark 
content $(s'd'u)$, but, respectively, symmetrized or 
antisymmetrized spin and $U-$spin wave functions. 
$W-$exchange  would not influence $\Xi'^0$. The other 
three states would be influenced, but differently. $n'$ 
can not go into any other of those states, while 
$\Lambda'$ and $\Sigma'^0$ mix to each other since the 
interaction~(7) can violate symmetrization of  $d'$ and 
$s'$.

Therefore, if the $U-$spin symmetry of strong and 
electromagnetic interactions were exact, $W-$exchange 
would eliminate degeneracy of states and separate 
definite combinations of them. One of the corresponding 
physical states, $\Sigma'^+$, would be unchanged by 
single $W-$exchange, whereas another, $p'$, would be 
changed. In particular, the mass of $p'$ would shift
from the unchanged mass of $\Sigma'^+$. From now on 
we can discuss problems related to Hara's theorem 
without invoking $U-$spin symmetry. Its  only role 
has been to make possible the mixing of $d$ and $s$ 
quarks.
   
Now we are ready to consider the photon emission vertex. 
Since the photon has been assumed from the beginning to 
be $U-$spin invariant, and $W-$exchange itself can not 
transform $\Sigma'^+$ to $p'$ or vice versa, they can not 
do it together as well. So photon emission (or absorption) 
can transform each of these two states either to itself or 
to heavier quark hadrons, but not to the other. Here we 
are interested in diagonal vertices. It is purely 
electromagnetic for $\Sigma'^+$ (remember that we 
account only for the lowest order, i.e. single 
$W-$exchange). The photonic vertex of $p'$ has an 
additional contribution due to $W-$exchange that 
violates $P-$ and $C-$parity. However we assume it 
to conserve combined $CP-$parity. Then the axial part 
of the vertex should be $C-$even and, therefore, can 
not contain any axial magnetic term (it would be axial, 
but $C-$odd, just as the usual magnetic term). 

This fact proves Hara's theorem specifically for 
$W-$exchange since it is just this additional 
vertex contribution that produces photonic 
transition between $\Sigma^+$ and $p$. Such an 
approach does not work for the transition between
$\Xi^-$ and $\Sigma^-$ since $W-$exchange is 
impossible for this pair, because of absence of 
$u-$quarks.

\section{Photonic vertex and gauge invariance} 

Now we discuss how  gauge invariance manifests itself  in 
the vertex  $p'p'\gamma$. Remember that we take into account 
the $W-$exchange contribution that violates space and charge 
parities separately. Therefore, the effective interaction 
takes the form
\begin{equation}
L_{\gamma} = e J'_{\mu} A^{\mu}
\end{equation}
with current $J'_{\mu}$ having both vector and axial parts.
Their matrix elements for $p'$ are
\begin{equation}
<p'| J'^V_{\mu} |p'> = \overline\psi_{p'}\left[f'_1(k^2)
\gamma_{\mu} + f'_2(k^2)\sigma_{\mu\nu}k^{\nu} + 
f'_3(k^2)k_{\mu}\right] \psi_{p'} ,
\end{equation}
\begin{equation}
<p'|J'^A_{\mu}|p'> = \overline\psi_{p'} 
\left[g'_1(k^2)\gamma_{\mu}\gamma_5 
+ g'_2(k^2) i\sigma_{\mu\nu}\gamma_5 k^{\nu} 
+ g'_3(k^2) i\gamma_5 k_{\mu}\right]\psi_{p'}.
\end{equation}
When $k \rightarrow 0$ only terms  $f'_1(0)\gamma_{\mu}$ 
and $g'_1(0)\gamma_{\mu}\gamma_5 $ can survive. Standard 
application of the gauge-invariance condition leads to 
the conclusion
\begin{equation} 
g'_1(0) = 0 ,
\end{equation} 
thus leaving us with only one term. Normalization  to 
the usual electric charge gives the further  relation 
\begin{equation} 
f'_1(0) = 1 . 
\end{equation}
These relations are 
quite usual and familiar, and should not raise any 
questions. However, explicit calculations in perturbation 
theory directly violate them both. To see this, one can repeat 
the calculations of ref.~\cite{KR}, applying them to diagonal 
transitions. The calculations are straightforward, and 
we will not describe them here. Instead we consider what 
is the reason for and meaning of such results.

Denote the "bare" propagator of $p'$ (i.e., with strong 
and electromagnetic interactions taken into account but 
without any weak interactions) with 4-momentum $q$ as 
$S_0 (q)$. It may be written as
\begin{equation}
S_0^{-1}(q) = a_0 (q^2) \hat q - M_0 (q^2) .
\end{equation} 
On the mass-shell, $q^2 = m_0^2 $. Near the
mass-shell
\begin{equation}
S_0 \approx ( \hat q - m_0 )^{-1},
\end{equation}
i.e., $a_0(m_0^2) = 1$ and $M_0(m_0^2) = m_0$.

Weak interactions (in particular, $W-$exchange) produce 
an additional self-energy part $\Sigma (q)$. So the total
$p'-$propagator $S (q)$ is determined by the expression     
\begin{equation}
S^{-1} (q) = S_0^{-1} (q) - \Sigma (q) .
\end{equation}
Analogously, we denote the "bare"  $p'p'\gamma-$vertex as 
$\Gamma^{(0)}_{\mu} (q_1,q_2,k)$ where $q_1$ and $q_2$ are 
the momenta of the initial and final $p'$ respectively, 
and $k$ is the photon momentum. $ \Gamma^{(0)}_{\mu}$ is 
purely vector. When both $q_1$ and $q_2$ are on the 
mass-shell it takes the standard form with Dirac and 
Pauli form-factors. Weak interactions produce an 
additional contribution $\delta \Gamma_{\mu}$ to the 
total $p'p'\gamma-$vertex.  

It is time now to recall that weak interactions (and 
$W-$exchange, in particular) violate parity. Therefore,
$\Sigma (q)$ may be written as 
\begin{equation}
\Sigma (q) = -a_+(q^2)\hat q - a_-(q^2)\hat q 
\gamma_5 + M_+(q^2) + M_-(q^2) i\gamma_5.
\end{equation} 
Similarly, $\delta\Gamma_{\mu}$ contains both vector 
and axial parts.

With weak interactions switched on, the mass-shell is 
determined by the physical mass $m$. Near it we may write
\begin{equation}
S^{-1} (q) \approx (a_0 + a_+)\hat q +
a_-\hat q\gamma_5  - (M_0 + M_+) - M_- i\gamma_5,
\end{equation}
where all functions are taken at $q^2 = m^2$. Therefore,
the Dirac equation for a free physical $p'$ (i.e., without 
any explicitly applied external field, but with complete 
account for the self-interaction) should be written as
\begin{equation}
[ (a_0 + a_+)\hat q + a_-\hat q\gamma_5 - (M_0 + M_+) -
M_- i\gamma_5 ] \psi_{p'} = 0 . 
\end{equation}
The physical mass $m$ is certainly related to parameters 
entering this equation, but we consider the relation 
somewhat later; for now we discuss the structure of the 
electromagnetic vertex.  

We emphasize here that the photonic vertex is not gauge 
invariant by itself. Instead, gauge invariance gives a 
general relation between propagator and vertex: 
\begin{equation}
\Gamma_{\mu}(q,q,0) = \frac{\partial S^{-1}(q)}
{\partial q^{\mu}}
\end{equation}
at $q^2 = m^2$. So we have
\begin{equation}
f'_1(0) = a_0(m^2) + a_+(m^2) ,\,\,\,\,\,
g'_1(0) = a_-(m^2) .  
\end{equation}
The same result appears from eq.~(18) for the minimal 
electromagnetic interaction. Note that generally $a_0(m^2) 
\neq 1$, since $a_0(m_0^2) =1$, and the strong interactions 
make $a_0(q^2)\neq $\,const (compare eqs.~(13),(14)); by the 
same reasoning $M_0(m^2)\neq m_0$. Thus, we see that the 
violation of the usual relations (11),(12) is directly 
traced to the non-canonical form of the Dirac equation (18)
and to the corresponding changes in applying gauge invariance. 
If parity were conserved, eq.~(11) would be satisfied, but 
eq.~(12) could nevertheless be violated. This situation is 
really well known, since eq.~(12) should work only after 
renormalization of both propagator and vertex. Eq.~(18) can 
also be transformed to the canonical form by a transformation 
of the wave function that may be considered as a generalized 
renormalization. Correspondingly, the propagator of $p'$ 
admits a transformation giving it the standard form near 
the physical mass-shell.

We define 
\begin{equation}
\psi' = Z^{-1/2} \psi ,\,\,\,\,\overline\psi' = 
\overline\psi\,\, \bar Z^{-1/2},
\,\,\,\,S'(q)= Z^{-1/2}S(q)\,\bar Z^{-1/2} .
\end{equation}
Here  $Z$\, is a constant matrix and $\bar Z = 
\gamma^0 Z^\dagger\gamma^0$. To arrive at the canonical 
form for the propagator and Dirac equation we take $Z$ 
as a product of three factors
\begin{equation}
Z =\frac{1}{[(a_0 + a_+)^2 - a_-^2]^{1/2}}\cdot 
\frac {(a_0 + a_+) - a_-\gamma_5}{[(a_0 + a_+)^2 - 
a_-^2]^{1/2}}\cdot\frac{M_0 + M_+ - M_- 
i\gamma_5}{[(M_0 + M_+)^2 + M_-^2]^{1/2}}\,\, .
\end{equation}
Correspondingly,
\begin{equation}
\bar Z =\frac{1}{[(a_0 + a_+)^2 - a_-^2]^{1/2}}\cdot 
\frac {(a_0 + a_+) + a_-\gamma_5}{[(a_0 + a_+)^2 - 
a_-^2]^{1/2}}\cdot\frac{M_0 + M_+ - M_- 
i\gamma_5}{[(M_0 + M_+)^2 + M_-^2]^{1/2}}\,\, .
\end{equation}
We consider the operator part (in square brackets) of 
eq.~(18) as $S'^{-1}(q)$ at $q^2 = m^2$ and transform 
it in accordance with eq.~(21). Then the above factors 
work as follows. The factor containing parameters $M$ 
takes $\gamma_5$ away from the mass terms of eq.~(18). 
The matrix factor with parameters $a$ does the same 
for the coefficient of $\hat q$. And the purely
numerical factor normalizes this coefficient to unity.
After that the Dirac equation for $\psi'$ takes its 
familiar form with the physical mass
\begin{equation}
m^2 = \frac{(M_0 + M_+)^2 + M_-^2}{(a_0 + a_+)^2 - 
a_-^2}\,\, .
\end{equation}
In accordance with eq.~(19) the vertex function should 
also transform as
\begin{equation}
\Gamma'_{\mu}(q_1,q_2,k) = \bar Z^{1/2}
\Gamma_{\mu}(q_1,q_2,k)Z^{1/2} .
\end{equation}
Consider how the matrix renormalization influences
the vertex. The factors of expressions (22),\,(23),
containing parameters $M$, mix to each other the vector 
and axial magnetic terms, as well as the induced scalar 
and pseudoscalar terms, and change their relative 
intensity. But the transformation may not remove any of 
them totally. The matrix factors with parameters $a$ 
mix vector and axial terms of the vertex and provide
the relation~(11). After that the remaining numerical 
part of the renormalization makes the relation~(12) be 
true as well. So the new vertex $\Gamma'_{\mu}$ has 
just the conventional limit as $k\rightarrow 0$ . Note
that $g'_1(k^2)$ at $k^2\neq 0$ may be non-vanishing
even after the total renormalization.

Note also that all the above arguments which should lead 
to the vanishing coefficient function $g'_2$ \,for the axial 
magnetic term in eq.~(10) can be applied only to the 
renormalized vertex $\Gamma'_{\mu}$ but not to $\Gamma_
{\mu}$. The reason is that the non-canonical form  
of the Dirac equation (18) generates a non-canonical expression
for the charge-conjugation transformation. Therefore, 
familiar notions on charge-conjugation behavior of various
vertices become distorted.   

Formally, the renormalization by the matrices (22),\,(23) is 
true only at
$$ |a_-| < |a_0 + a_+| ,$$
which is satisfied since $a_0 \sim \cal O$\,$(1)$, 
while $a_+$ and $a_-$ are induced by weak interactions. 
Eq.~(24) shows that the opposite case would lead to the 
tachyon ($m^2 < 0$).

The matrix renormalization can also be formulated in terms 
of the unmixed states $p$ and $\Sigma^+$. In such a form
the renormalizing matrix would have even more complicated 
structure to account for both parity violation and particle
mixing.

Now we are ready to understand the origin and meaning of 
the results obtained by Kamal and Riazuddin~\cite{KR}. They 
studied amplitudes for the transition $\Sigma^+\rightarrow 
p$ which is closely related to our amplitudes for the 
diagonal transition $p'\rightarrow p'$. So we will discuss 
their results in terms of our amplitudes. 

Consider $<p'|\Gamma_{\mu} e^{\mu}|p'>$ where the vector 
$e^{\mu}$ may be thought of as a photon polarization 
vector. Note that we do not make any additional 
renormalization (just as in ref.~\cite{KR}). If we study 
only terms of zeroth order in $k$ (i.e. take the limit 
$k\rightarrow0$, again as in ref.~\cite{KR}) we obtain
\begin{equation}
<p'|\Gamma_{\mu}(q,q,0) e^{\mu}|p'> =
\overline\psi_{p'}(q)[f'_1(0)\hat e + g'_1(0)\hat e 
\gamma_5]|\psi_{p'}(q).
\end{equation}   
It is simplest to calculate the right-hand side in the 
rest frame. Then, due to properties of the Dirac matrices, 
only the time component $e^0$ contributes to the first term 
in r.h.s., while the second term contains 
$\boldmath{(e\cdot\sigma)}$ with only space components 
contributing. If we take $e^0 = 0$ (again, as in 
ref.~\cite{KR};\, it looks only natural for the photon 
polarization vector) then we see an axial contribution 
without any vector one. 

It is just this result that was promoted ~\cite{KR} 
as contradicting Hara's theorem, according to which 
one would expect to find only a vector contribution, 
without an axial one. We see, however, that the result 
has really no relation to terms for which we should 
apply Hara's theorem. Those terms may appear only 
in calculations which completely account for the 
first order in $k$. Moreover, the correct relation 
between the physical vector and axial magnetic terms 
arises only after matrix renormalization~(25) of the 
full photonic vertex. This means that only after such
renormalization one may compare results of perturbative 
calculations and Hara's theorem predictions.

Note, ironically, that in the case of initially exact 
$U-$spin symmetry the decay $\Sigma^+ \rightarrow p\gamma$
would be impossible at all (again, to lowest order in
$W-$exchange). Indeed, as we have seen, real physical states 
in that case would be not $\Sigma^+$ and $p$, but $\Sigma'^+$
and $p'$  which are incapable of transforming into one another 
by photon emission, though  their differing masses allow the 
decay kinematically.

It is instructive, nevertheless, to see how the transition 
amplitude  $\Sigma^+ \rightarrow p\gamma$ would look in the
case of the initially exact $U-$symmetry. To do this we 
reverse the mixing transformation and express $\Sigma^+$ 
and $p$ through  $\Sigma'^+$ and $p'$. Then we may substitute 
the vertices $p'p'\gamma$ and $\Sigma'^+ \Sigma'^+\gamma$  
(which is pure electromagnetic; remember that transitions
between $\Sigma'^+$ and $p'$ are absent) and obtain the 
desired amplitude. If we apply the inverse mixing 
transformation to the non-renormalized $p'$ we arrive 
at the amplitude having a structure that violates canonical 
expectations of both gauge invariance and Hara's theorem. 
It contains non-vanishing vector and axial terms at 
$k^2 = 0$ as well as axial magnetic term. Only if we apply 
the transformation to the renormalized $p'$ ($\Sigma'^+$ 
is not influenced by $W-$exchange) the resulting amplitude 
looks as expected. Hence, renormalization of $p'$ touches
both $p$ and $\Sigma^+$. 

This demonstrates importance of matrix renormalization even 
for the case of exact $U-$spin symmetry. If symmetry violation is 
more intensive than the influence of $W-$exchange, the 
states $p'$ and $\Sigma'^+$ do not arise. But then we 
should renormalize both $p$ and $\Sigma^+$ taking into 
account their transitions to each other through the 
$W-$exchange.

\section{Conclusion and discussion}

Let us briefly summarize the above discussion. Here we
have  reconsidered Hara's theorem and explicitly
formulated its assumptions. Results of the theorem 
are in a very close relation to the well-known 
properties of usual $\beta-$decay, as was noted 
earlier by Gaillard~\cite{MG}. Assumptions of 
the theorem are rather simple and clear-cut. When 
they are satisfied the theorem is surely true. And 
they are satisfied in many approaches used in the 
literature. 

Of course, one of the assumptions for Hara's
theorem, $U-$spin symmetry, is violated in 
nature and in calculations. However, in many 
applications its violation may be considered 
as small. That a similar possibility does not 
work for WRD's\, has caused a long-standing 
problem of their description which is still 
unsolved.

We have demonstrated, in particular, that  for
$W-$exchange Hara's theorem should also be true. 
More detailed study of $W-$exchange contributions 
shows that an effect stated some years ago as 
manifesting violation of Hara's theorem for 
$W-$exchange at the quark level does not really 
necessitate such violation. Instead, it can be 
explained as revealing insufficiency of standard 
purely numerical renormalization  in perturbation 
theory for weak interactions. If parity is 
violated the fermion renormalization "constants"
should be taken as combination of the unit matrix
with $\gamma_5$.  

The "violation" of gauge invariance described in 
the preceding sections may look strange and even
mysterious. But its reason is really quite clear.
Conservation of the current $J'_{\mu}$ means 
that its matrix elements (9) and (10) taken over
real physical states should vanish when being
multiplied by $k^{\mu}$. But each r.h.s. of eqs.
(9),\,(10) consists of three elements. The square 
brackets contain the effective vertex with various 
matrix terms and corresponding coefficient functions,
Two other elements are wave functions, initial and 
final. When expanding into perturbation series, all 
the three elements should be expanded. Meanwhile, 
standard application of gauge invariance implicitly 
assumes that wave functions have simple free 
structure and need not be expanded. It is true,
but only after renormalization. 

This familiar and trivial fact becomes not so trivial, 
when the mixing of various parities or even various 
particles is involved. Admixture of $p$ to $\Sigma^+$
(and vice versa) induces admixture of "bare" 
electromagnetic vertices to the transition vertex.
"Ideologically" the situation is reminiscent of the 
so-called pole approach to WRD, but the formulas may 
look unlike.
 
The present consideration shows that previous
calculations of WRD amplitudes may need 
some revision since they have not taken into 
account necessity of a non-standard 
renormalization procedure. The possibility of 
particle mixing in weak interactions makes this 
procedure even more complicated.
 
In this regard, we would like to emphasize 
the large role of particle mixing in the 
present discussion and its possible role in 
future calculations. Such mixing is essentially 
nonperturbative, in the sense that while its 
coefficients depend on symmetry properties of 
the perturbation, they are independent of its 
intensity. Moreover, the mixing may complicate 
the apparent consequences of gauge invariance. 
Therefore, an accurate account for the mixing 
might open new ways to describe large 
symmetry-violation effects observed in WRD. 

One more lesson which may also be of general
character concerns properties of renormalization 
constants. It is clear that they need to be 
Lorentz-invariant. But if parity is violated 
there are no arguments why the renormalization
of fermionic propagators and vertices could not use
the matrix $\gamma_5$\,, instead of being purely 
numerical. And as the above discussion, together with
calculations of Kamal and Riazuddin~\cite{KR}, shows,
at least in that particular case one should apply 
matrix renormalization (i.e., include $\gamma_5$ 
into renormalization constants) to have canonical 
expressions for the Dirac equation, gauge and 
charge-conjugation transformations, and so on. 
The same question arises, therefore, for radiative 
corrections in any weak processes.   

\section*{Acknowledgements}

I acknowledge the hospitality of Fermilab Theory Division
where this work was finished and prepared for publication.
It is a pleasure to thank J.Lach and P.Zenczykowski for 
many useful discussions which, in particular, stimulated 
my attention to the result of Kamal and Riazuddin. I also 
thank J.Lach, K.Ellis and D.Richards for reading the 
manuscript.

\end{document}